# INSULATOR-CONDUCTOR TRANSITION: A BRIEF THEORETICAL REVIEW

**M. Cattani and M. C. Salvadori**
<mcsalvadori@if.usp.br> and <mcattani@if.usp.br>
Institute of Physics, University of São Paulo, C.P. 66318, CEP 05315-970 São Paulo, Brazil

**F. S. Teixeira**
Electronic Systems Engineering Department, Polytechnic School, Av. Prof. Luciano Gualberto
Trav. 3-158, University of São Paulo, CEP 05508-900 São Paulo, Brazil

**Abstract.** The electrical conductivity of disordered insulator-conductor composites have been studied for more than thirty years. In spite of this some properties of dc bulk conductivity of composites still remain incompletely understood. We present a brief review of the most significant theories that have been proposed to study the critical insulator-conductor transition comparing their predictions with many experimental results.

## (1) Introduction.

The transport properties of disordered insulator-conductor composites have been studied for more than thirty years. In spite of this some phenomena still remain incompletely understood [1]. According to many proposed theories of transport in isotropic percolating materials [1-5] the dc bulk conductivity σ of a composite, near the critical conductor-insulator transition, is given by the scaling power law

$$\sigma \propto (p - p_c)^t \quad . \qquad (1.1)$$

In the above equation p is the probability of occupation of a site in resistor network by a conducting element, $p_c$ is the *critical probability* for bond percolation, below which the composite has zero conductivity (or more precisely the conductivity of the insulating phase) and t is the *critical exponent*. The above expression holds true in the critical region $p - p_c \ll 1$ in which critical fluctuations extend over distances much larger than the characteristic size of the constituents of the insulating-conductor composite.

Sometimes, instead of (1.1), the conductivity σ of random resistor networks is written in the form,

$$\sigma \approx \sigma_o (x - x_c)^t \qquad (1.2),$$

where $\sigma_o$ is a proportionality constant, x the volume concentration of the conducting phase and $x_c$ is the *critical concentration* below which the composite has zero conductivity. Sometimes the conductance σ is indicated by g.

It is important to note that in the *critical percolation phenomena* we can distinguish three conduction regimes: (1) *metallic or conductor* ( $x > x_c$), (2) *transition* ($x \approx x_c$) and (3) *insulator* or *dielectric* ($x < x_c$). In the metallic regime the composite behaves like a dirty metal; the resistivity is relatively low and the temperature coefficient of the resistivity is positive.

A vast class of disordered conducting – insulating compounds has been analyzed in the last thirty years. In Fig.1 is shown [1] a collection of different 99 measured values of critical exponent t and the corresponding critical threshold concentration $x_c$ for various disordered conductor-insulator composites. From these results we verify that the critical parameters t and $x_c$ vary in the ranges, $1.5 \leq t \leq 11$ and $0.05 \leq x_c \leq 0.5$. These various composites include carbon–black–polymer systems, oxide–based thick film resistors and other metal–inorganic and metal–organic insulator composites.

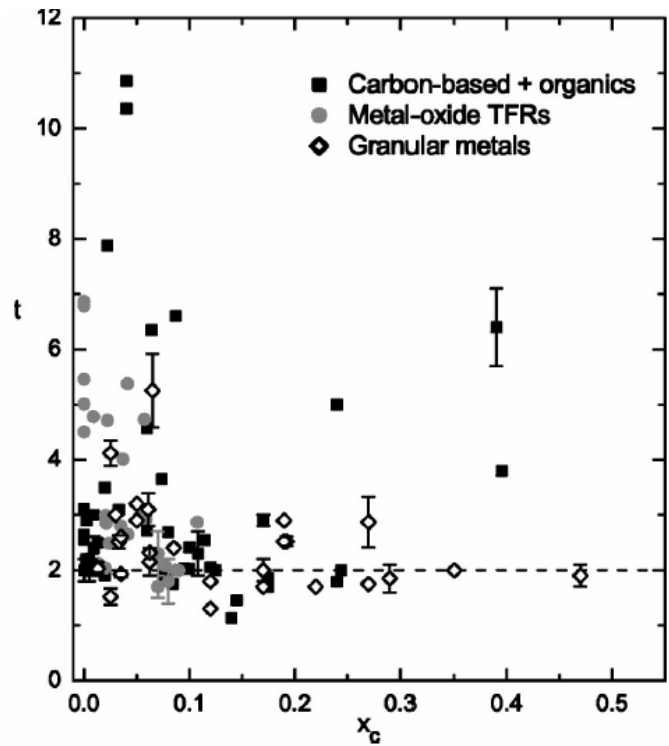

Fig.1. Collection of critical exponent values t and corresponding critical threshold concentration $x_c$ for various disordered insulator-conductor composites.

Roughly we can say that there are essentially three different approaches to explain the percolation and conduction in random resistor networks: *Cluster Theory*, *Resistor Network Theory* and *Tunneling-Percolation Theory*. These cases will be analyzed in the following Sections.

**(2) Cluster Theories.**

In these models the percolation and transport phenomena in the composites are calculated taking into account the formation of *clusters* in d dimensional lattices. These clusters are composed by neighbor insulator and conducting "elements" which occupy the sites in lattices [4]. The conducting "elements" are usually represented by black balls (or occupied sites) and white balls (or empty sites) are the insulating ones. Since the formation of clusters is a *random* process, *statistics* is the basic mathematical tool used to investigate the percolation and conductance processes. Each site of a lattice is randomly occupied with probability p and the empty site with probability (1 – p) and clusters are groups of neighboring occupied sites

By series expansions and numerical calculations the percolation and the conductance σ have been calculated for various lattices [4] with dimensions d = 2, 3,…,7. The 1-dim case and the *Bethe lattice* (or Cayley tree) are solved exactly. These calculations have shown that the percolation is a critical phenomenon and that the conductance σ at the threshold obeys the scaling law given by the Eq.(1.1). In Table 1 are shown [4] the percolation thresholds $p_c$ for various lattices and dimensions d assuming two different models: *site percolation* and *bond percolation*. We show here only the results for the 2D square lattice (square) and for the 3D cases: simple cubic (SC),
BBC and FCC. For all these cases the critical exponent t was found to be t = 2.00.

| *Site percolation* | *Bond percolation* |
|---|---|
| $p_c$ = 0.593(square) | $p_c$ = 0.500 |
| $p_c$ = 0.312 (SC) | $p_c$ =0.249 (SC) |
| $p_c$ = 0.246 (BCC) | $p_c$ = 0.180 (BCC) |
| $p_c$ = 0.198 (FCC) | $p_c$= 0.119 (FCC) |

**Table 1.** The percolation thresholds $p_c$ for the 2D square lattice (square) and for the 3D cases SC, BCC and FCC assuming two different models: *site percolation* and *bond percolation*.

Somewhat different values for $p_c$ and t have been obtained for SC lattice with the cluster theory applying new renormalization group insights to the percolation problem [6-8]. Kirpatrick [6] have found $p_c = 0.312$ and $t = 1.8 \pm 0.05$; Sur et al.[7] have found $p_c = 0.311$ and $t = 1.6 \pm 0.1$; Levinshtein et al.[8] have found $t = 1.69 \pm 0.03$.

**(3) Resistor−Network Theory.**

In this case it is supposed that we known the detailed spatial [2,3] arrangement of the conducting and nonconducting materials in a composite system. If all dimensions of the conducting regions are large with respect to the electronic mean free paths, a local conductivity $\sigma(r)$ can be defined by the bulk value of the conductivity for the material at the point r. The random arrangement of the material modifies the conductivity of a sample in several ways. The problem is to calculate the effective conductivity of the composite material taking into account the statistical distribution of $\sigma(r)$´s in the material. Given $\sigma(r)$ this is done solving the usual equations of electrostatics,

$$\mathbf{j}(r) = \sigma(r)\, \mathrm{grad}V(r)$$

$$\mathrm{div}\, \mathbf{j}(r) = 0, \qquad (3.1)$$

where $\mathbf{j}(r)$ and $V(r)$ denote the local current and voltage, respectively, which result when a field is applied across the sample. Equations (3.1) may be solved to any desired accuracy using a finite difference approximation [2,3,5,9]. A convenient discrete model is obtained substituting the composite continuous medium by a resistor network. In this case we have resistors connecting the nodes of a lattice. Indicating by $V_i$ the voltages at the nodes of each network and by $\sigma_{ij}$ the *random* conductance of the link between adjacent nodes *i* an *j*, the condition that all currents into the node *i* cancel is

$$\sum_j \sigma_{ij}(V_i - V_j) = 0. \qquad (3.2)$$

This equation is just the discrete form of the condition $\mathrm{div}\, \mathbf{j}(r) = 0$ and corresponds to the Kirchhoff current law.

The effective conductance of the medium $\sigma_m$, or simply $\sigma$, is obtained solving Eqs.(3.2) taking into account the effect of the $\sigma_{ij}$ random values.

Kirkpatrick [2,3] has proposed three different *percolation models* to calculate $\sigma_m$ which depend of somewhat different hypothesis about the statistical distribution and spatial correlation of the $\sigma_{ij}$ : (a)*bond percolation model*,(b) *site percolation model* and (c)*correlated bond percolation model*. He has assumed that the random resistor networks $\sigma_{ij}$ obey a binary distribution, that is, they have only two values, $\sigma_1$ and $\sigma_2$, with probabilities $p_1 = p$ and $p_2 = 1- p$, respectively. It was verified that these simple binary networks are interesting in their own right since clearly they exhibit a percolation threshold and they can be related to real composite materials.

In these conditions Kirpatrick [2, 3] has solved numerically by the Monte Carlo method the Eqs.(3.2) to calculate the effective conductivities of large regular 2D square lattice (square) and 3D simple cubic (SC) lattice networks. It was assumed the simplest case of $\sigma_1 = 1$ and $\sigma_2 = 0$. He has shown that in the threshold $\sigma$ obeys the scaling law $\sigma \sim (p - p_c)^t$. With the *bond percolation model* he obtained $p_c = 0.5$ and $t = 2$ for the square lattice and $p_c = 0.25$ and $t = 1.6 \pm 0.1$ for the SC lattice. With the *correlated bond percolation model*, $p_c = 0.103$ and $t = 1.5 \pm 0.2$ for the SC lattice. With the *site percolation model*, $p_c = 1/3$ and $t = 1.5 \pm 0.2$ for the SC lattice.

Several authors [10-12] have solved numerically Eqs.(3.2) taking into account multifractal properties of the current distribution of 3D random resistor network at the percolation threshold. Assuming that $p_c \approx 0.25$ they have found $t \approx 2.00$ for the SC lattice.

Kirkpatrick [2] has re-examined and generalized the old *effective-medium theory* of conduction in mixtures to treat the resistor networks. He has shown (see Appendix A) that if the $\sigma_{ij}$ are distributed according to some distribution function $p(\sigma)$ the effective conductivity $\sigma_m$ is the solution of the equation,

$$\int d\sigma \, p(\sigma) \frac{(\sigma_m - \sigma)}{\sigma + (z/2 - 1)\sigma_m} = 0, \qquad (3.3)$$

where z is the number of bonds at each node of the network. This equation permits us to calculate analytically the effective conductance $\sigma_m$.

Taking into account a binary conductance distribution $p(\sigma)$, $\sigma_1$ and $\sigma_2$, defined by $p(\sigma) = p\, \delta(\sigma - \sigma_1) + (1-p)\, \delta(\sigma - \sigma_2)$ we can show [2], using Eq.(5), that $\sigma_m$ is given by:

$$\sigma_m = [(zp/2-1)\sigma_1 + [z(1-p)/2-1]\sigma_2]/(z-2) + \\ + [\{(zp/2-1)\sigma_1 + [z(1-p)/2-1]\sigma_2\}^2 + 2(z-2)\sigma_1\sigma_2]^{1/2}/(z-2). \qquad (3.4)$$

The $\sigma_m$ predictions obtained with Eq.(3.3) are surprisingly accurate [2,3] since they are in fair agreement with the numerical Monte Carlo calculations. At the threshold it was found $p_c = \frac{1}{2}$ and $t = 2$ for the square lattice ($z = 4$) and $p_c = 1/3$ and $t = 3/2$ for SC lattice ($z = 6$).

**(3) Tunneling-Percolation Theory**

As was shown above, the predicted values for the critical exponent t obtained with the *cluster*, *resistor-network* and the *effective-medium theories* are in the range $2.0 \geq t > 1.5$. However, a large number of experimental results in a vast class of conducting-insulating compounds [1,13] have shown that t varies in the range $12 > t > 1.5$. This discrepancy between the numerous experimental results and the available theoretical predictions is still an unresolved issue. A major difficulty [13] in the comparison of the theoretical predictions with the experimental results appears to be the lack of experimental information about the local structural geometry of the composites. To overcome this discrepancy and to explain the experimental results it was proposed a new approach named *Tunneling-Percolation Theory* (TPT).

According to the TPT it is assumed that the conducting particles are embedded in an insulating matrix and that the electric transport between the particles is due to *percolation* and *tunneling*. The coexistence of tunneling and *percolation* has been recently settled by experiments probing the electrical connectivity of various disordered systems [1, 14]. The low-field electrical [15] tunneling conductivity $\sigma_L$ of many disordered materials has a temperature dependence that can be expressed in the form $\sigma_L \sim \exp(-b/T^p)$. The value $p = 1/4$, found in many of the amorphous semiconductors and semiconducting glasses has been predicted by Mott [16] using a model of hopping conductivity between localized states. There are evidences [14] that $p = \frac{1}{2}$ for granular metals, consisting of fine metallic particles dispersed in a dielectric matrix, and that this behavior can be explained by a structural effect.

When the grain charge effects can be neglected with respect with the tunneling processes it is assumed [1] that the interparticle tunneling conductance g is given by,

$$\sigma = \sigma_o \exp[-2(r-\Phi)/\xi] \qquad (4.1),$$

where $\sigma_o$ is a constant, $\xi$ is the tunneling factor which is of order of few nm, r the distance between the centers of two neighboring spheres of diameter $\Phi$ ( $r \geq \Phi$ for impenetrable particles).

Taking into account the tunneling conductance $\sigma$ given by Eq.(4.1) and following the calculations done by P.M.Kogut and J.Straley [17], Balberg [18] and Grimaldi and Balberg [13] it can be shown (see Appendix B) that the *percolation-tunneling conductance* $\Sigma$ near the threshold, is written as

$$\Sigma \sim (p - p_c)^T \qquad (4.2),$$

where the new critical exponent T is given by (see Appendix A),

$$T = t \quad \text{if} \quad \alpha < 0$$
$$T = t + 1/(1-\alpha) \quad \text{if} \quad 0 < \alpha < 1, \quad (4.3)$$

$\alpha = 1 - \xi/2(a - \Phi)$ and a is the mean interparticle distance.

**APPENDIX A - Generalized effective medium theory.**

We show here how to obtain Eq.(3.3) following the calculations developed by Kirkpatrick [3].

Let us take an infinite rectangular network composed by resistors with random-valued $\sigma_{ij}$ resistivities. Indicating by $V_i$ the voltages at the nodes of each network and by $\sigma_{ij}$ the *random* conductance of the link between adjacent nodes $i$ and $j$, the condition that all currents into the node $i$ cancel is

$$\sum_j \sigma_{ij} (V_i - V_j) = 0 \quad (A.1).$$

This equation corresponds to the Kirchhoff current law.

Let us indicate by $\sigma_m$ the average value of the conductances $\sigma_{ij}$ that is, $\sigma_m = <\sigma_{ij}>$. We imagine now a network made up of equal conductances, $\sigma_m$, connecting the nearest neighbors on the cubic mesh. In this *uniform net* the adjacent nodes $i$ and $j$ are separated by a potential difference $V_m$. This would be an *effective average network* or simply an *effective medium*. In this homogeneous medium the average value of the local electric field is equal to zero and, consequently, the average value of the differences of potential between any two nodes would be also equal to zero.

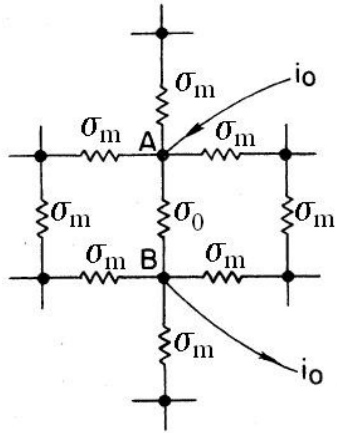

**Fig.A.1.** Construction used to show how to calculate the voltage induced across one conductance, $\sigma_o$, surrounded by an uniform medium.

Now we replace in this uniform medium, between the points A and B, the conductance $\sigma_m$ by as shown in Fig. A.1. If a current $i_o$ is introduced at the node A and extracted at B (see Fig. A.1), the uniform solution fails to satisfy current conservation at A and B. If $V_m$ is the potential difference between A and B, that is, $V_{AB} = V_m$ to correct this the current $i_o$ is chosen to obey the condition

$$i_o = V_m (\sigma_m - \sigma_o). \qquad (A.2)$$

The extra voltage, $V_o$, induced by $i_o$ between A and B, can be calculated if we know the conductance $\Sigma_{AB}^*$ of the medium between A and B, excluding $\sigma_o$ (see Fig.A.2). So, the current $i_o$ created by the difference of potential $V_{AB} = V_o$ would be given by,

$$i_o = V_o (\Sigma_{AB}^* + \sigma_o). \qquad (A.3)$$

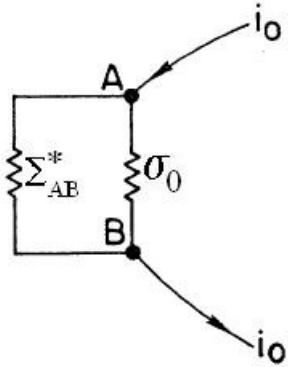

**Fig.A.2.** Construction equivalent to that seen in Fig.A.1 where is shown the effective conductance $\Sigma_{AB}^*$ of the medium between A and B.

A symmetry reasoning is useful: express the current distribution in Fig.A.1 with $\sigma_{AB} = \sigma_m$ as the sum of two contributions, a current $i_o$ introduced at A and extracted at a very large distance in all directions, and an equal current, introduced at infinity and extracted at B. In each case the current flowing through each of the z equivalent bonds at the point where the current enters is $i_o/z$, so a total current of $2i_o/z$ flows through the AB bond. This means that $i_m = 2i_o/z$. Consequently, putting $\Sigma_{AB} = \Sigma_{AB}^* + \sigma_m$ we get

$$V_m \Sigma_{AB} = i_o = z\, i_m/2 \qquad (A.4)$$

Since $V_m \sigma_m = i_m$ (A.4) can be written as $(i_m/\sigma_m) \Sigma_{AB} = i_o = z\, i_m/2$, giving

$$\Sigma_{AB} = (z\, \sigma_m/2). \qquad (A.5)$$

This implies that

$$\Sigma_{AB}^* = \Sigma_{AB} - \sigma_m = (z/2 - 1)\sigma_m . \qquad (A.6)$$

Using (A.3) we obtain $V_o = i_o/[\sigma_o + (z/2 - 1)\sigma_m]$ and, finally, remembering that $i_o = V_m(\sigma_m - \sigma_o)$, according to (A.2), we have:

$$V_0 = \frac{V_m(\sigma_m - \sigma_0)}{\sigma_0 + (z/2 - 1)\sigma_m} . \qquad (A.7)$$

If the (random) bond values $\sigma_{ij}$ are distributed according to a probability distribution $p(\sigma)$ (continuous or discrete), the requirement that the average of $V_o$ value is given by $<V_o> = 0$, permit us to write (A.7) as

$$\int d\sigma\, p(\sigma) \frac{(\sigma_m - \sigma)}{\sigma + (z/2 - 1)\sigma_m} = 0, \qquad (A.8)$$

from which $\sigma_m$ can be calculated, according to (3.3).

**APPENDIX B – Tunneling-percolation conductance.**

Kogut and Straley [17], considering an infinite random lattice network and using the *generalized effective medium theory* of Kirkpatrick[2,3], assumed that the random conductivities $\sigma$ obey the distribution $g(\sigma)$ given by,

$$g(\sigma) = p\, h(\sigma) + (1-p)\, \delta(\sigma) \qquad (B.1)$$

Now, they imagined that the lattice is formed by isolators and by a large number of poor conductors so that $h(\sigma) \sim \sigma^{-\alpha}$ for small $\sigma$, with $0 < \alpha < 1$. With this hypothesis they were able obtain critical exponents larger than the "universal value", that is, $t > 2$. It is easy to verify numerically using (3.4) that with these hypothesis t can assume values larger than 2.

Let us show how to get (4.3) following the calculations presented in reference [1] taking

$$\sigma = \sigma_o \exp[-2(r - \Phi)/\xi] \qquad (B.2)$$

they have obtained the distribution function $h(\sigma)$ using the following equation

$$h(\sigma) = \int_\Phi^\infty dr\, P(r)\, \delta\{\sigma - \sigma_o \exp[-2(r-\Phi)/\xi]\} \qquad (B.3).$$

In this equation P(r) is the distribution function of adjacent intersphere distances r. Assuming that the normalized distribution of interparticle distances P(r) is well represented by [1]

$$P(r) = \exp[-(r-\Phi)/(a-\Phi)]/(a-\Phi) \qquad (B.4)$$

we see that h(σ) defined by Eq.(B.3) becomes,

$$h(\sigma) = \frac{1-\alpha}{\sigma_0}(\sigma/\sigma_0)^{-\alpha} , \qquad (B.5)$$

where $\alpha = 1 - \xi/2(a-\Phi)$. This equation (B.5) agrees with the original hypothesis of Kogut and Straley [17], that is, $h(\sigma) \sim (1-\alpha)\sigma^{-\alpha}$ .

**Acknowledgements.** This work was supported by the Fundação de Amparo a Pesquisa do Estado de São Paulo (FAPESP) and the Conselho Nacional de Desenvolvimento Científico e Tecnológico (CNPq), Brazil.